\documentclass[useAMS,usenatbib]{mn2e}
\usepackage{graphicx,fleqn,fix2col,rotating}
\DeclareGraphicsExtensions{.eps,.ps,.eps.gz,.ps.gz,.eps.Z}
\title[The photometric period in ES Ceti]{The photometric period in ES Ceti}

\author[C.M.~Copperwheat et al.]{C.M.~Copperwheat$^{1}$, T.R.~Marsh$^{1}$,  V.S.~Dhillon$^{2}$, S.P.~Littlefair$^{2}$,\newauthor P.A.~Woudt$^{3}$, B.~Warner$^{3}$, J.~Patterson$^{4}$, D.~Steeghs$^{1}$,\newauthor J.~Kemp$^{4}$, E.~Armstrong$^{4}$ and R.~Rea$^{5}$\\\\
$^{1}$ Department of Physics, University of Warwick, Coventry, CV4 7AL, UK\\
$^{2}$ Department of Physics and Astronomy, University of Sheffield, Sheffield, S3 7RH, UK\\
$^{3}$ Astrophysics, Cosmology and Gravity Centre, Department of Astronomy,\\ University of Cape Town, Private Bag X3, Rondebosch 7701, South Africa\\
$^{4}$ Department of Astronomy, Columbia University, 550 West 120th Street, New York, NY10027, USA\\
$^{5}$ CBA (Nelson), 8 Regent Lane, Richmond, Nelson, New Zealand\\
}

\date{Received: }

\begin{document}

\newcommand{\dg} {^{\circ}}
\outer\def\gtae {$\buildrel {\lower3pt\hbox{$>$}} \over
{\lower2pt\hbox{$\sim$}} $}
\outer\def\ltae {$\buildrel {\lower3pt\hbox{$<$}} \over
{\lower2pt\hbox{$\sim$}} $}
\newcommand{\ergscm} {erg s$^{-1}$ cm$^{-2}$}
\newcommand{\ergss} {erg s$^{-1}$}
\newcommand{\ergsd} {erg s$^{-1}$ $d^{2}_{100}$}
\newcommand{\pcmsq} {cm$^{-2}$}
\newcommand{\ros} {{\it ROSAT}}
\newcommand{\xmm} {\mbox{{\it XMM-Newton}}}
\newcommand{\exo} {{\it EXOSAT}}
\newcommand{\sax} {{\it BeppoSAX}}
\newcommand{\chandra} {{\it Chandra}}
\newcommand{\hst} {{\it HST}}
\newcommand{\subaru} {{\it Subaru}}
\def\rchi{{${\chi}_{\nu}^{2}$}}
\newcommand{\Msun} {$M_{\odot}$}
\newcommand{\Mwd} {$M_{wd}$}
\newcommand{\Mbh} {$M_{\bullet}$}
\newcommand{\Lsun} {$L_{\odot}$}
\newcommand{\Rsun} {$R_{\odot}$}
\newcommand{\Zsun} {$Z_{\odot}$}
\def\Mdot{\hbox{$\dot M$}}
\def\mdot{\hbox{$\dot m$}}
\def\mincir{\raise -2.truept\hbox{\rlap{\hbox{$\sim$}}\raise5.truept
\hbox{$<$}\ }}
\def\magcir{\raise -4.truept\hbox{\rlap{\hbox{$\sim$}}\raise5.truept
\hbox{$>$}\ }}
\newcommand{\mnras} {MNRAS}
\newcommand{\aap} {A\&A}
\newcommand{\apj} {ApJ}
\newcommand{\apjl} {ApJL}
\newcommand{\apjs} {ApJS}
\newcommand{\aj} {AJ}
\newcommand{\pasp} {PASP}
\newcommand{\apss} {Ap\&SS}
\newcommand{\araa} {ARAA}
\newcommand{\nat} {Nature}
\newcommand{\pasj} {PASJ}
\newcommand{\sovast} {Soviet Astron.}
\maketitle

\begin{abstract} 
We present ULTRACAM photometry of ES Cet, an ultracompact binary with a $620$s orbital period. The mass transfer in systems such as this one is thought to be driven by gravitational radiation, which causes the binary to evolve to longer periods since the semi-degenerate donor star expands in size as it loses mass. We supplement these ULTRACAM+WHT data with observations made with smaller telescopes around the world over a nine year baseline. All of the observations show variation on the orbital period, and by timing this variation we track the period evolution of this system. We do not detect any significant departure from a linear ephemeris, implying a donor star that is of small mass and close to a fully degenerate state. This finding favours the double white dwarf formation channel for this AM CVn star. An alternative explanation is that the system is in the relatively short-lived phase in which the mass transfer rate climbs towards its long-term value.
\end{abstract}

\section{INTRODUCTION}  

Ultracompact binaries with periods of the order of tens of minutes or less have attracted much attention in recent years. These systems consist of a primary white dwarf with a companion star that is also at least partially degenerate. Close double-degenerate binaries are one of the proposed progenitor populations of Type Ia supernovae \citep{Kotak08,Distefano10,Gilfanov10},  as well as the recently proposed sub-luminous `.Ia' supernovae \citep{Bildsten07,Kasliwal10}. They are of interest from a binary formation and evolution point of view, with the short periods implying at least one common envelope phase in the history of the binary, and the chemical composition suggesting helium white dwarfs, helium stars or cataclysmic variables (CVs) with evolved secondaries as possible progenitors \citep{Nelemans01,Nelemans10}. The mass transfer in these systems is thought to be driven by angular momentum loss as a result of gravitational radiation. These sources are predicted to be among the strongest gravitational wave sources in the sky \citep{Nelemans04}, and are the only class of binary with examples already known which will be detectable by the gravitational wave observatory LISA \citep{Stroeer06,Roelofs07}. 

The two systems with the shortest known periods are HM Cnc ($324$s, \citealt{Roelofs10}) and V407 Vul ($569$s, \citealt{Haberl95}). The exact nature of these systems is unclear: the two leading models are the unipolar induction (UI) model \citep{Wu02} and the direct-impact accretion model \citep{Marsh02,Roelofs10}. There are also 23 known systems with longer periods ranging from $620$ to $3906$s (the AM Canum Venaticorum stars, AM CVns; see \citealt{Solheim10} for a recent review). The natures of these systems are better established: the spectra show an absence of hydrogen and the presence of helium lines, many of which are the double-peaked emission lines characteristic of sources accreting via a disc \citep{Marsh99,MoralesRueda03}. These systems are clearly accreting, and are the helium equivalent of the cataclysmic variable (CV) stars.

Gravitational radiation has a huge influence on these systems, driving the evolution and determining the orbital period distribution, luminosities and numbers. Measurements of the time derivative of the orbital period $P$ for V407 Vul and HM Cnc have shown the periods in these systems to be decreasing \citep{Hakala03, Hakala04, Ramsay05, Strohmayer05}. This is consistent with the UI model, and contrary to what might be expected for accretion. In an accreting double-degenerate system, the gravitational radiation should drive the binary towards longer orbital periods. However, if the mass transfer in an accreting double-degenerate system is significantly lower than its equilibrium value, due to either some non-secular process \citep{Marsh05} or the binary being in its mass transfer turn-on phase \citep{Willems05,DAntona06,Deloye06}, then a decreasing binary period is possible. The counter-argument to this is that such phases are expected to be short, although \citet{Deloye06} calculated that the early contact phase can last much longer ($10^3$ - $10^6$ yr) than originally thought.

To date, no period derivative has been measured in any of the longer period AM CVn stars. Since these are unambiguously accreting binaries, then we would expect to detect an increasing period in these objects. Additionally, it is not certain that gravitational radiation is the dominant angular momentum loss mechanism for the ultra-compact binary stars: some other mechanism, such as the magnetic braking observed in cataclysmic variables, could be operating to drive the evolution at a higher rate than assumed in current models \citep{Farmer10}. These astrophysical issues will in the future be important for the use of these systems for the verification of LISA as they can make the difference between detectability or not \citep{Stroeer06}. 

In order to address this issue we have conducted a long-term timing study of the AM CVn star ES Ceti \citep{Warner02}. This system shows a $620$s optical modulation which is complex in structure and varies on a night-by-night basis \citep{Espaillat05}. Spectroscopic observations have confirmed that this is the orbital period, and also show the double-peaked helium emission lines which imply the presence of an accretion disc in this system (Steeghs, in prep). ES Cet is the ideal subject for a timing study, since it is the shortest period ultracompact binary population after V407 Vul and HM Cnc. It potentially connects these two systems to the rest of the AM CVn population: the period of ES Cet is only $51$s greater than V407 Vul, so ES Cet may also be in the mass transfer turn-on phase. If however ES Cet has already evolved on to the long-term and stable AM CVn path of lengthening period, then we would expect the period evolution to be the most rapid of all the systems, since gravitational radiation is strongest when masses are large and periods short.

\section{OBSERVATIONS}
\label{sec:obs}

\subsection{WHT/ULTRACAM}

\begin{figure*}
\centering
\includegraphics[angle=270,width=1.0\textwidth]{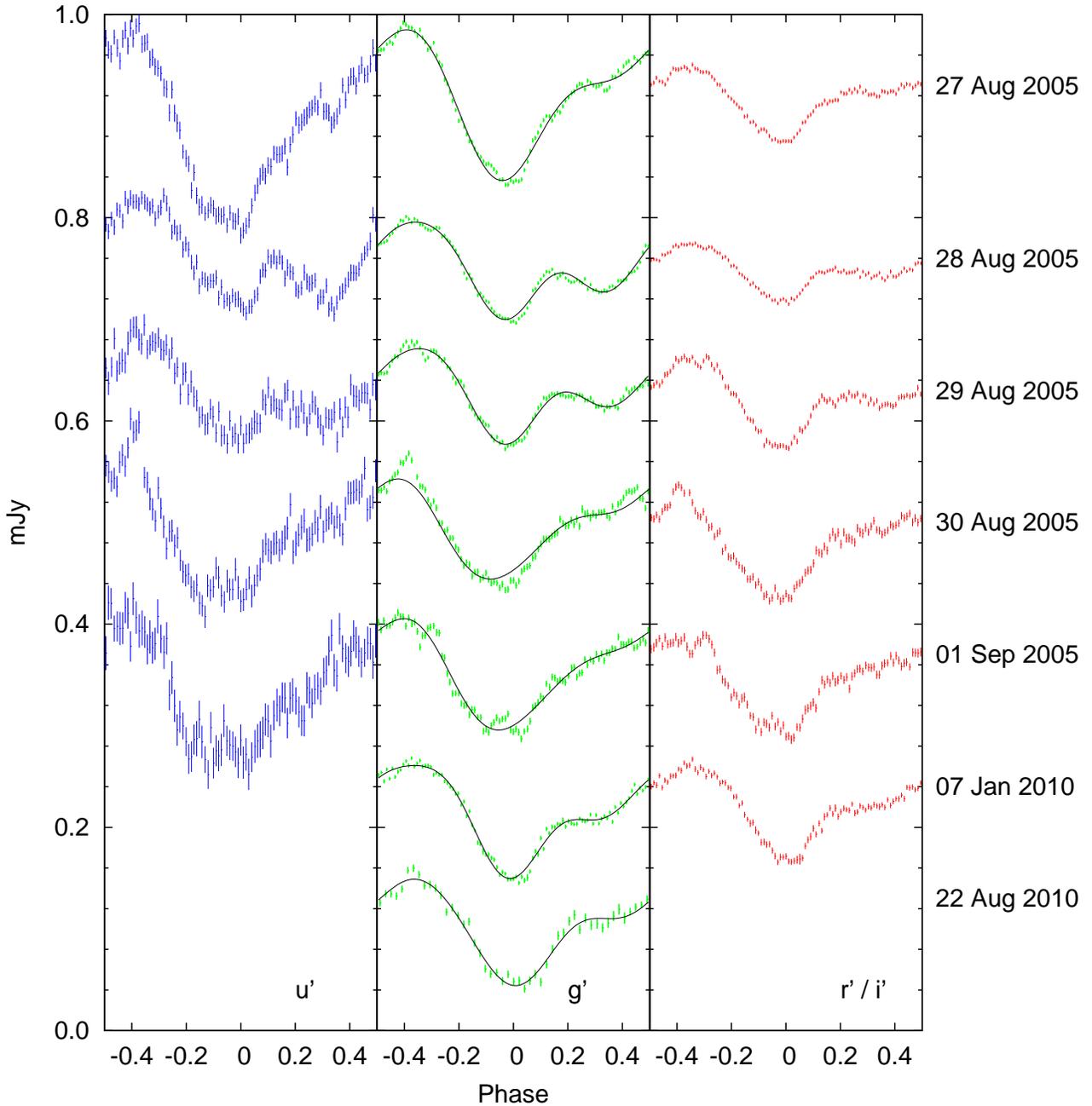}
\hfill
\caption{Phase folded ULTRACAM and ACAM light curves. Due to the significant variations in the light curves from night-to-night, we phase fold the data on a night by night basis. The $u'$-band light curves are plotted in the left column, the $g'$-band light curves in the middle, and the right column contains either the $i'$-band (nights 1 and 2) or $r'$-band (nights 3 to 6) light curves. We omit the $u'$-band light curve for night 6, which was particularly affected by poor weather conditions. A flux offset is applied to each light curve for clarity. For the $g'$-band data we overplot the fits as detailed in Section \ref{sec:analysis}, consisting of three sinusoids at the fundamental orbital frequency and its first and second harmonics. } \label{fig:lightcurves} \end{figure*}

We observed ES~Cet with the high speed CCD camera ULTRACAM \citep{Dhillon07} mounted on the $4.2$m William Herschel Telescope (WHT) in $2005$ and $2010$. ULTRACAM is a triple beam camera and all observations were made using the SDSS $u'$, $g'$ and $r'$ or $i'$ filters. The CCDs were windowed in order to obtain exposure times of $2$ -- $4$s, depending on conditions. The dead time between exposures for ULTRACAM is $\sim$$25$ms. In 2005 we observed ES Cet five times over the period 27 Aug -- 01 Sep $2005$. Conditions were good, with seeing $< 1$'' and good transparency. We observed ES Cet once more on the night of 07 Jan $2010$. These observations were affected by poorer seeing and some cloud. 

All of these data were reduced with aperture photometry using the ULTRACAM pipeline software. We follow the same steps for reduction and flux calibration as detailed in our previous ULTRACAM publications (e.g. \citealt{Copperwheat10a,Copperwheat10b}).

\subsection{WHT/ACAM}

We made a further observation of ES~Cet in excellent conditions on the night of 22 August 2010 with the WHT using the permanently-mounted imaging camera ACAM \citep{Benn08}. We used a 5 second exposure time and windowed the chip to reduce the readout time between exposures to $\sim$$0.8$s. These data were reduced in the same way as the ULTRACAM data.

\subsection{SAAO and CBA observations}

We supplemented our WHT observations with additional data collected with smaller telescopes over a longer time period. We observed ES Cet with the University of Cape Town CCD photometer \citep{Odonogue95}, mounted on the $1.9$m telescope at the South African Astronomical Observatory (SAAO). This instrument was used in frame transfer mode and the observations were made in white light in order to maximize the count rate. Typical exposure times vary from $10$ -- $20$s. These data were reduced using the {\sc DoPhot} program \citep{Schechter93,Odonogue95}. We discarded any data in which conditions, the exposure time or the length of the observation means that the photometric period is not readily apparent. This left us with $38$ nights of observation obtained between $18$ Oct $2001$ and $9$ Oct $2009$. The length of each observation varied, but was typically between $5$ and $10$ orbital cycles.

We obtained additional nights of observation with the Center for Backyard Astrophysics (CBA) telescope network, a global network of small telescopes dedicated to photometry of cataclysmic variables\footnote{More details concerning the CBA can be found at http://cba.phys.columbia.edu}. Typical exposure times for these observations were $\sim$$20$s. As with the SAAO data, we discarded any observations in which the photometric period is not obvious. This leaves us with $21$ observations obtained between $18$ Jan $2002$ and $1$ Jan $2010$.

\section{ULTRACAM LIGHT CURVES}
\label{sec:lightcurves}

We measured the flux from ES Cet in 2005 using our ULTRACAM data, and find the mean magnitudes to be $15.94$, $16.28$, $16.63$ and $17.23$ in $u'$, $g'$, $r'$ and $i'$ respectively. We found the object to be slightly fainter in 2010, with mean magnitudes of $16.19$, $16.35$ and $16.73$  in $u'$, $g'$ and $r'$. The amplitude of the light curve varies between bands, but is $0.11$ - $0.18$ on average.

Using the ephemeris given in Section \ref{sec:analysis} we phase-folded our ULTRACAM and ACAM data and plot the results in Figure \ref{fig:lightcurves}. We do not combine data from multiple nights, since \citet{Espaillat05} showed evidence for some variation in the shape of the light curve on different nights. We also observe these changes in our data: while each light curve is roughly sinusoidal with the minimum light consistently at the phase of zero, on 28 and 29 August 2005 we see a secondary minimum at a phase of $\sim$$0.35$, which is not present in the other three light curves obtained in 2005. The two 2010 light curves may be intermediate between the two states observed in 2005. The timescale over which the light curve shape can change is very short: an examination of the unfolded light curves shows significant changes over the course of a few orbital cycles. The cause of this variation is unclear, but it is most likely associated with the accretion disc.

\section{ORBITAL TIMINGS}
\label{sec:analysis}

\begin{table}
  \caption{Our minimum-light timings for ES Cet, using the sine function fit described in Section \ref{sec:analysis}. We determine the uncertainties by fitting a large number of data sets, resampled from the original data using the bootstrap method \citep{Efron79,Efron93}. For each timing we indicate if it was obtained by SAAO (S), CBA (C), ULTRACAM (U) or ACAM (A).}
  \label{tab:timings}
  \begin{center}
    \begin{tabular}{llllll}
&Cycle&BMJD&&Cycle&BMJD\\
\hline    
S&0	    &	52200.98063(3)	&	C&57688	&	52615.08678(6)	\\
S&134	&	52201.94270(3)	&	C&60619	&	52636.12651(3)	\\
S&274	&	52202.94762(3)	&	S&60994	&	52638.81849(10)	\\
S&433	&	52204.08908(3)	&	S&83192	&	52798.16412(4)	\\
C&12698	&	52292.13155(4)	&	S&93497	&	52872.13709(5)	\\
C&12838	&	52293.13636(3)	&	S&93779	&	52874.16180(6)	\\
C&16596	&	52320.11289(4)	&	S&97951	&	52904.10985(5)	\\
S&16831	&	52321.79975(7)	&	S&99892	&	52918.04294(5)	\\
S&16968	&	52322.78333(5)	&	S&134317	&	53165.15846(4)	\\
S&17107	&	52323.78128(5)	&	S&134875	&	53169.16426(7)	\\
S&43614	&	52514.05836(2)	&	S&146279	&	53251.02627(6)	\\
S&43745	&	52514.99882(4)	&	S&146847	&	53255.10348(3)	\\
S&44162	&	52517.99198(3)	&	S&167703	&	53404.81578(7)	\\
C&47844	&	52544.42274(3)	&	S&167841	&	53405.80629(6)	\\
C&47982	&	52545.41327(2)	&	S&167981	&	53406.81140(8)	\\
C&48121	&	52546.41115(2)	&	U&196318	&	53610.22484(1)	\\
C&48392	&	52548.35663(4)	&	U&196454	&	53611.20136(1)	\\
C&48675	&	52550.38796(2)	&	U&196595	&	53612.21349(1)	\\
S&48763	&	52551.01975(3)	&	U&197006	&	53615.16357(2)	\\
S&49035	&	52552.97227(4)	&	S&245612	&	53964.07579(4)	\\
S&49051	&	52553.08725(4)	&	S&245762	&	53965.15266(7)	\\
C&49096	&	52553.41030(3)	&	C&259428	&	54063.25214(5)	\\
S&49158	&	52553.85520(5)	&	C&265117	&	54104.09019(8)	\\
S&49181	&	52554.02031(3)	&	C&315130	&	54463.10209(8)	\\
C&49235	&	52554.40809(4)	&	C&317361	&	54479.11720(6)	\\
S&49302	&	52554.88571(38)	&	C&319869	&	54497.12051(5)	\\
S&49441	&	52555.88667(4)	&	C&359186	&	54779.35267(4)	\\
C&49509	&	52556.37485(2)	&	S&405672	&	55113.04683(4)	\\
S&52360	&	52576.84022(5)	&	C&417121	&	55195.23200(8)	\\
S&52535	&	52578.09646(8)	&	C&417387	&	55197.14168(5)	\\
S&54316	&	52590.88107(6)	&	U&418324	&	55203.86763(2)	\\
S&56127	&	52603.88493(28)	&	A&449996	&	55431.22123(2)	\\
S&56682	&	52607.86516(5)	&				

   \end{tabular}
  \end{center}
\end{table}


In order to measure the photometric period in ES Cet we choose to fit a simple function to each night of data. We found that the variation in ES Cet is fit well on any given night by the sum of three sinusoids of the form $f(t) = a_n \cos(2\pi n (t - t_0 - t_n) / P)$ where $n$ is $1$, $2$ or $3$, hence these sinusoids correspond to the fundamental orbital frequency and the first and second harmonics. $P$ is the orbital period and $t_0$ is the time of minimum light. We fitted this function to each night of ULTRACAM $g'$-band data (in which the signal-to-noise is highest). These fits are over-plotted on the $g'$-band data in Figure \ref{fig:lightcurves}. We first set $t_0$ to $0$ and $P$ to $1$ and fit the function to the phase folded light curve in order to determine $a_n$ and $t_n$. We then determined $t_0$ from the unfolded light curve by fixing $a_n$ and $t_n$ and fitting the function to these data. The process of fitting the data is complicated by the variable shape of the light curve, which we discussed in Section \ref{sec:lightcurves}. This particularly complicates the task of fitting the CBA/SAAO observations, which have a much lower cadence than our WHT observations. It is therefore difficult to fit most of the observations with the same degree of precision with which we fit the WHT observations. 

Furthermore, we find that our determination of the light curve minimum $t_0$ is perturbed by the changing light curve shape. This is apparent in the $2005$ ULTRACAM timings. These were collected over the course of a few successive nights, and so when the linear ephemeris is subtracted they should have similar $O-C$ values. We find the $O-C$ values to form two groups with a significant offset of around $20$s (with a formal error on the individual timings of $\sim$$1$s). These two groups correspond to the two distinct light curve shapes observed in Figure \ref{fig:lightcurves}. In an attempt to characterise this effect with the aim of subtracting it from our light curves, we fitted each night separately and examined the variation in the model parameters with the $O-C$ excess from a linear ephemeris. We found first of all that the model parameters can take a continuous range of values between the two extreme states we observed with ULTRACAM, and we observed a scatter of up to $\sim$$25$s in the timings. Some component of this scatter is doubtless due to flickering, the stochastic variation observed in all accreting systems; but a large component is due to the gross changes in the light curve. We observed a loose correlation between the $O-C$ values and the amplitude and phase of the first harmonic. We attempted to improve our $O-C$ residuals by adding an offset determined from these correlations, but since the correlation is fairly weak our timings were not improved to any significant degree. The limitation here is most likely the difficulty in fitting the majority of our observations with precision, due to the low time resolution and the gross variation of the light curve on timescales of a few cycles. Since any improvement is very small, we chose not to apply any correction of this nature to our final determinations of the orbital timings.

For consistency we obtain our final timings by applying the same model fit to each night, allowing $t_0$ to vary but fixing $a_n$ and $t_n$. In practise we found it does not matter which night of ULTRACAM data we use for these $a_n$ and $t_n$ values, but we choose night $6$ since the light curve shape in this case is intermediate between the two extremes. We did try phase-folding and fitting each night separately, and then using the fit for each night to determine the time of minimum light for that individual night, but this did not improve the timings to any significant degree and in some cases made things worse due to the uncertainty of the fits. We obtain $65$ timings in total, which we list in Table \ref{tab:timings}. Our final set of timings is in good agreement (RMS $\sim$$6$s) with the measurements of \citet{Espaillat05} for the cases where we share data. We convert all timings to the barycentric dynamical timescale, correcting for light travel times.

Finally, using these timings, we calculate an ephemeris of \\

\noindent $BMJD(TDB) = 52200.980575(6) +  0.00717837598(3)E$\\\\
for the minimum-light point of the orbital cycle, using a linear least-squares fit. The data points in Figure \ref{fig:timings_mdot} show the residuals after this linear fit is subtracted, and demonstrates that there is no significant departure from linearity over the baseline of our observations.

\section{DISCUSSION}
\label{sec:discussion}

In this section we examine the period change in ES Cet over the baseline of our observations. In Section \ref{sec:periodchange} we discuss the period changes and subsequent departure from a linear ephemeris in AM CVn systems as a result of gravitational radiation. We then compare these models to our ES Cet timings in Section \ref{sec:periodev}.

\subsection{Period changes in accreting binaries}
\label{sec:periodchange}

The fractional period change in an accreting binary system can be shown to be 
\begin{equation}
\frac{\dot{P}}{P} = 3 \left( \frac{\dot{J}}{J} - (1-q)
\frac{\dot{M}_2}{M_2} \right), \label{eq:accrete}
\end{equation}
\citep{Marsh05} assuming conservative mass transfer, and where $q = M_2/M_1$,  $M_1$ and $M_2$ are the masses of the accretor and donor respectively, and \Mdot$_2$ is the mass transfer rate. The corresponding quadratic time shift in the times of light curve minima is given by
\begin{equation}
\Delta T = {1 \over 2} \Bigl({\dot P \over P}\Bigr) t_B^2,\label{eq:shift}
\end{equation}
where $t_B$ is the baseline of our observations. Assuming that the spin of the accretor is strongly coupled to the orbit (see \citealt{Marsh04}), the $\dot J$ term in Equation \ref{eq:accrete} is simply due to gravitational radiation, where
\begin{equation}
{\dot{J}_\mathrm{GR} \over J_\mathrm{orb}} = - \frac{32}{5}
\frac{G^3}{c^5} \frac{M_1 M_2 M}{a^4},\label{eq:gr}
\end{equation}
\citep{Landau75}, where $a$ is the orbital separation. Since $\dot J / J < 0$ in the absence of mass transfer, we would expect a decreasing period as the two components spiral in. In an accreting system the \Mdot$_2$ \ term in Equation \ref{eq:accrete} can compensate for $\dot J$ (since \Mdot$_2$ \ $< 0$) and for the expected mass transfer rates in AM CVn stars it is larger, and so we expect a lengthening period in these systems once it has settled into its long-term, stable state. It can be shown that 
\begin{equation}
{\dot{M_2}\over M_2} =  {{\dot{J} / J} \over {1 + (\zeta_2 - \zeta_{RL})/2 - q}}\label{eq:mdot}
\end{equation}
\citep{Marsh04}, where 
\begin{eqnarray}
\zeta_2 &=& \frac{d \log R_2}{d \log M_2},
\end{eqnarray}
and
\begin{eqnarray}
\zeta_{RL} &=& \frac{d \log (R_L/a)}{d \log M_2}
\end{eqnarray}
describe the response of the donor and the Roche lobe to mass transfer. $\zeta_{RL}$ is taken to be $1/3$ using the small $q$ approximation of \citet{Paczynski71}. Since the response of a degenerate star is to expand on mass loss, $\zeta_2$ is $< 0$. A value of $\zeta_2 = -0.28$ is appropriate for a fully degenerate donor, but as the degree of degeneracy of the donor is decreased, $\zeta_2$ approaches $0$ \citep{Deloye05}.

\subsection{The period evolution in ES Cet}
\label{sec:periodev}

In this section we compare our measurements of the orbital timings with the predicted time shift over our baseline. We see in Section \ref {sec:periodchange} that the predicted time shift depends on the masses of the donor star and the accretor. We begin in Section \ref{sec:masses} by selecting various appropriate values for these unknown parameters. We then compare our timings with the expectations for these parameters in Sections \ref{sec:accretion} and \ref{sec:gr}. We consider first the case in which there is mass transfer between the two components. We also consider the detached case, in which there is no mass transfer and the period decreases at the rate given by Equation \ref{eq:accrete} with $\dot{M}_2 = 0$. However given that the spectral observations imply accretion, we would expect the first case to be a better fit to our data.

\subsubsection{The component masses}
\label{sec:masses}

There are three formation paths proposed for AM CVns. All three paths are consistent with a donor that is partially degenerate to some degree, and measurements of the component masses in other AM CVn systems have confirmed the donors to be partially degenerate \citep{Roelofs06,Roelofs07a,Copperwheat10b}. The `white dwarf channel' \citep{Nelemans01} suggests detached close double white dwarfs which are brought into contact as a result of angular momentum loss due to gravitational wave radiation (GWR). \citet{Nelemans01} used a fully-degenerate, zero-temperature donor in their formulation, but \citet{Deloye05} argued that the donors could be semi-degenerate depending on the contact time of the binary. The second scenario is the `evolved CV channel' \citep{Podsi03} which suggests the progenitors of AM CVns are CVs with evolved secondaries. The donor in this channel is initially non-degenerate and hydrogen-rich, but becomes degenerate and helium-rich (but still with a few per cent hydrogen) during its evolution before Roche lobe overflow. The third scenario is the `helium star channel' \citep{Iben91}, in which the donor had been a helium core burning star before coming into contact with its Roche lobe. For the white dwarf and evolved CV channels the donor mass is somewhat arbitrary, since it depends on factors such as the specific entropy of the donor at time of contact \citep{Deloye05}. We therefore chose to examine two cases: the first in which the donor is fully degenerate, and the second in which it has a mass consistent with a helium star progenitor. The helium star channel implies a very massive donor star, and the expected mass for the other two channels will lie between these two extremes.

If we assume a fully degenerate donor, we find $M_2$ to be $0.062$\Msun \citep{Deloye05}. This is the lowest possible mass for the donor in this system. We chose to combine this with the lowest likely accretor mass $M_1$, since the case in which both $M_1$ and $M_2$ are at their lowest values produces the minimum period change in both the accreting and non-accreting formulations. For this we assumed the system is accreting via a disc, rather than through the direct impact of the accretion stream onto the white dwarf primary. We believe accretion via a disc is a reasonable assumption, since the spectroscopic data do feature the double peaked lines which are characteristic of an accretion disc, although the direct-impact scenario cannot be ruled out entirely, since even in this case the primary Roche lobe might be filled with diffuse, hot gas which could produce a `disc-like' signature in the spectra \citep{Roelofs10}. The presence of an accretion disc implies a minimum mass for the accretor (for a given donor mass) simply via geometric arguments, since the accretion stream follows a ballistic trajectory. For a donor mass of $0.062$\Msun, the minimum possible mass for the accretor is $0.44$\Msun. This accretor mass is quite low, so we also considered a higher accretor mass of $\sim$$0.6$\Msun, which is more typical of field white dwarfs. Since the donor is fully degenerate, we set $\zeta_2 = -0.28$.

For the partially degenerate case, we use the \citet{Nelemans01} fit to the He-star M-R track of \citet{Tutukov89} to determine the donor mass. Figure 1 of \citet{Deloye05} shows that the predicted donor mass for this scenario is $\sim$$0.26$\Msun. We again assume the accretor mass to be at the minimum for accretion via a disc, which for this donor mass is $0.69$\Msun. The value of $\zeta_2$ in this case is not clear but will lie somewhere between $0$ and the fully-degenerate value of $-0.28$. Given that the donor star we consider here is considerably less degenerate than the $M_2 = 0.062$\Msun \ case we would expect $\zeta_2$ to be closer to $0$, but we consider both of these two extremes.

\subsubsection{Period evolution with mass transfer}
\label{sec:accretion}

\begin{figure*}
\centering
\includegraphics[angle=270,width=0.8\textwidth]{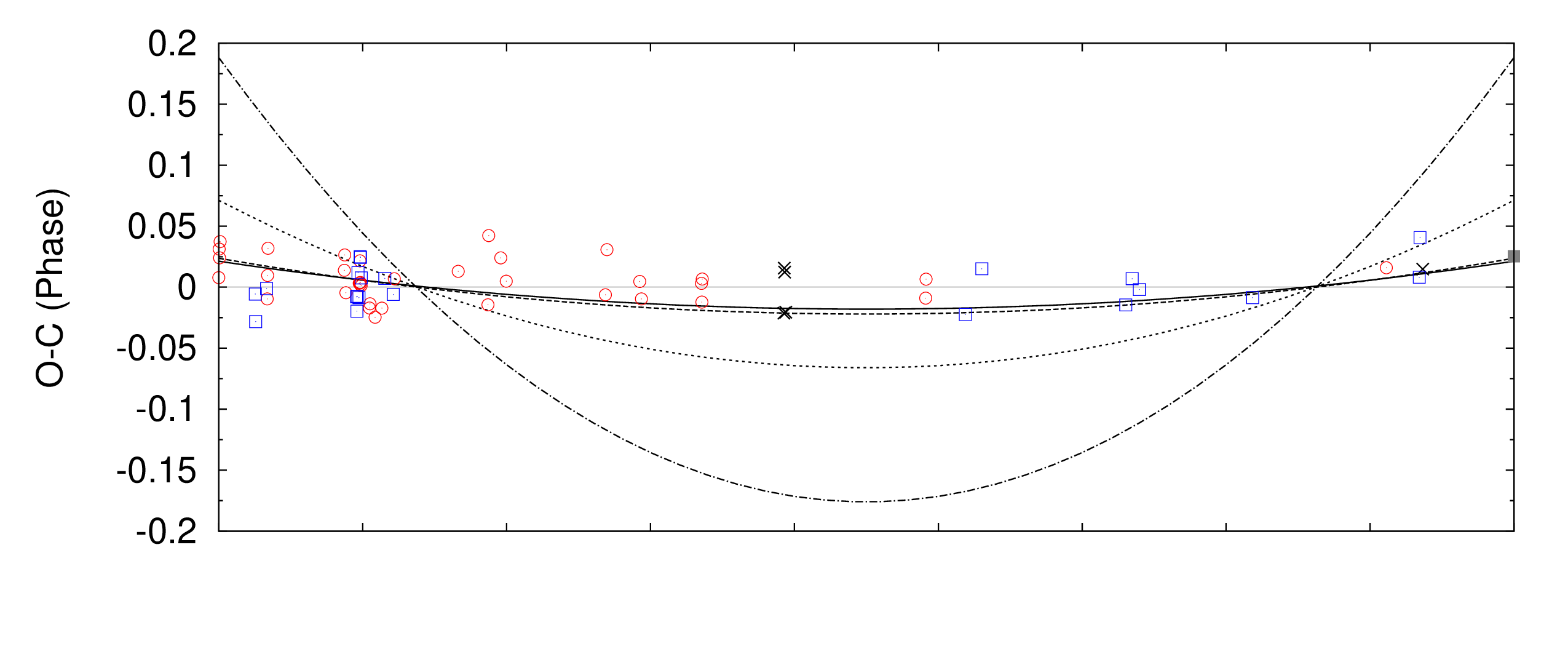}\vspace{-10mm}
\includegraphics[angle=270,width=0.8\textwidth]{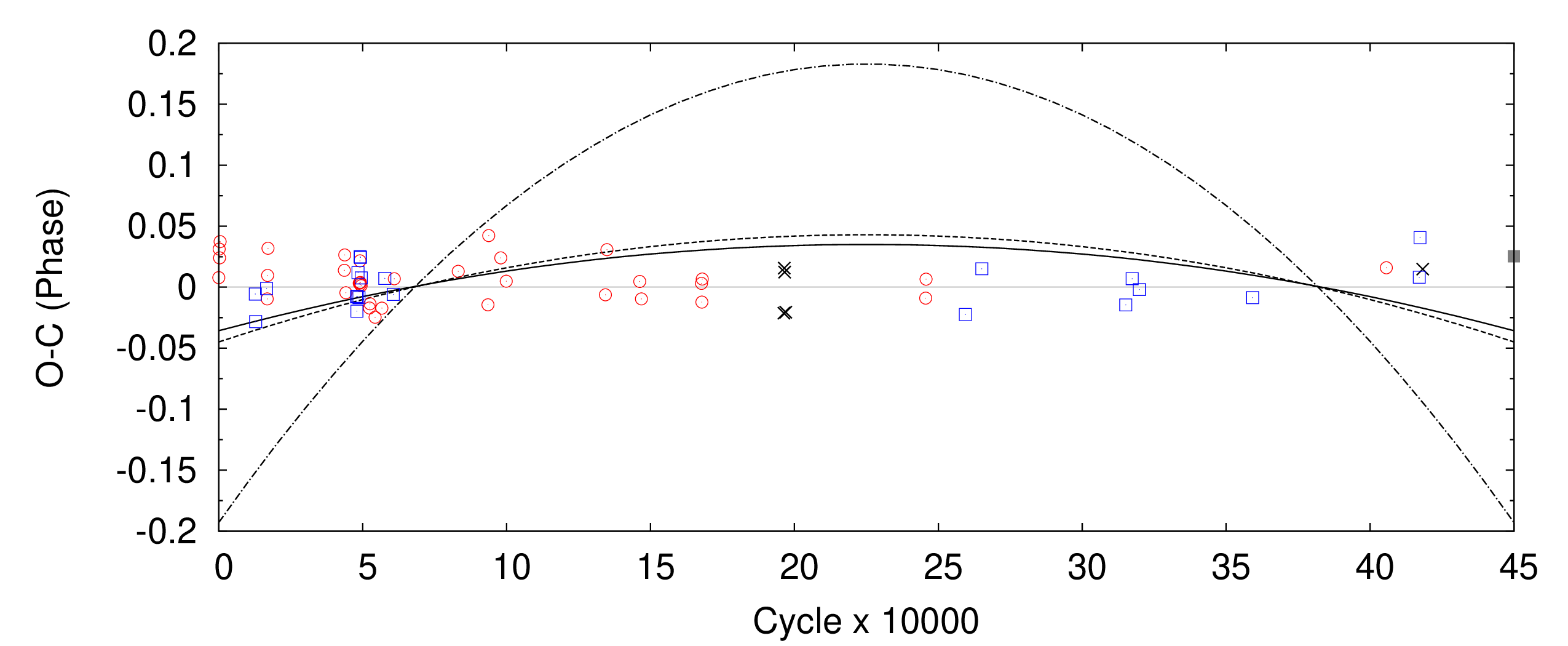}\vspace{0mm}
\caption{{\bf Top panel:} Minimum-light timings for our complete dataset. We plot the period excess $(O-C)$ in terms of phase against the orbital cycle. ULTRACAM observations are denoted by the black crosses, SAAO observations by the unfilled red circles, CBA observations by the unfilled blue squares, and the ACAM observation by the filled grey square.  The four curved lines indicate the expected period change when we include mass transfer at the secular rate, for different combinations of accretor and donor mass. The four combinations are {\bf (i)} solid line,  $M_1 = 0.44$\Msun, $M_2 = 0.062$\Msun, $\zeta_2 = -0.28$. This is the minimum period change, assuming accretion via a disc and a fully degenerate donor. {\bf (ii)} dashed line,  $M_1 = 0.60$\Msun, $M_2 = 0.062$\Msun, $\zeta_2 = -0.28$. This is a variation on the minimum case in which we assume a more likely accretor mass. {\bf (iii)} dot-dashed line,  $M_1 = 0.69$\Msun, $M_2 = 0.26$\Msun, $\zeta_2 = -0.28$. Here we assume a partially degenerate donor using the M-R relation for the He-star formation channel \citep{Nelemans01}, and the corresponding minimum accretor mass. We use the same value of $\zeta_2$ as for the fully degenerate models. This is unlikely to be correct, and so in {\bf (iv}, dotted line) we set $\zeta_2 = 0$, keeping all other parameters the same. The true value of $\zeta_2$ will be between these two extremes, but is likely to be closest to $0$. {\bf Bottom panel:} As with the top panel, except here we assume the secondary star is detached from its Roche lobe, and so the binary evolution proceeds due to gravitational radiation in the absence of mass transfer. In this detached case models (iii) and (iv) give the same prediction, since they are only distinguished from each other by the response of the donor to mass loss.} \label{fig:timings_mdot} \end{figure*}

We plot in Figure \ref{fig:timings_mdot} the deviation from a linear ephemeris for our orbital timings in terms of phase, using the linear ephemeris given in Section \ref{sec:analysis}. If we examine the data points first of all, we see a large scatter (as discussed in Section \ref{sec:analysis}) is apparent in these timings (in particular, note the separation between the two groups of 2005 ULTRACAM timings, at $\sim$$200,000$ cycles). However, aside from this random variation, there is no obvious departure from a linear ephemeris over the baseline of our observations. There might be a very small increase in period over our baseline (manifested by a slightly positive $O-C$ value at either extreme of the plot and a slightly negative value in the middle), however any such trend does not appear to be very significant when compared to the noise.

The curved lines in the top panel of Figure \ref{fig:timings_mdot} show the predicted departure from a linear ephemeris with $M_1 = 0.44$\Msun, $M_2 = 0.062$\Msun (solid line); $M_1 = 0.6$\Msun, $M_2 = 0.062$\Msun (dashed line) and $M_1 = 0.69$\Msun, $M_2 = 0.26$\Msun (the dot-dashed and dotted lines, for a $\zeta_2$ of $-0.28$ and $0$ respectively). All four predict an increasing period as a result of the mass transfer. These lines have been vertically offset to optimise the fit to the data. We see that in the cases where we assume a fully degenerate donor the predicted time shift is very small, and the amplitude is less than the uncertainty in our data points. A fully degenerate donor is therefore consistent with our timings. Increasing the mass of the accretor has little effect on these amplitudes. Conversely, the predicted time shift when a donor mass of $M_2 = 0.26$\Msun \ is used is much larger. When we set $\zeta_2$ to $-0.28$ it is very large, and clearly precluded by our measurements. As $\zeta_2$ is increased the response of the donor to mass loss is decreased, and so the expected time shift over our baseline is also decreased. The exact value of $\zeta_2$ for this semi-degenerate donor is likely close to (but not greater than) $0$. The dotted line in Figure \ref{fig:timings_mdot} shows that even when we set $\zeta_2 = 0$ the predicted deviation from linearity is not consistent with our timings if we assume a massive donor.

We have shown that the He-star formation channel produces a donor that is too massive to be consistent with our timings. If we examine the evolutionary tracks given in \citet{Podsi03} for the evolved CV channel we find that, while the donor mass is dependent on initial conditions, a reasonable assumption for the donor mass is $0.11$ -- $0.12$\Msun, which also predicts a deviation from linearity which is inconsistent with our timings. We conclude therefore that the double white dwarf channel is the most likely formation path for the low donor mass we infer.

An alternative explanation is that ES Cet is in the initial, turn-on phase before the long AM CVn phase of lengthening period, and so the accretion rate in ES Cet is currently below the stable rate which would be expected if it were in that phase. The decreasing periods of HM Cnc and V407 Vul have been explained by invoking an accretion model in which the donor has a thin hydrogen envelope, and it is the accretion of this envelope which leads to the shortening period \citep{DAntona06}. At some point this envelope will be depleted and the transfer of helium will begin, leading to a lengthening period along the `traditional' evolutionary track. As the binary transitions from one state to the other the fractional period change must drop significantly. Since we find ES Cet to be consistent with a $\dot P$ of $0$ it could now exist in this intermediate regime, linking HM Cnc and V407 Vul with the longer period AM CVns. 

Finally, we note that one assumption we have made is that all of the angular momentum contained in the mass gained by the accretor from the donor is fed back into the orbit. Alternatively, some fraction of the disc accretion may act to spin up the accretor. However, we find that this only increases the expected $\dot P$, and so does not improve the fit of any of our models to the observed timings. If we consider the extreme case, in which none of the angular momentum is fed back to the orbit, Equation \ref{eq:mdot} is modified to 
\begin{equation}
{\dot{M_2}\over M_2} =  {{\dot{J} / J} \over {1 + (\zeta_2 - \zeta_{RL})/2 - q} -\sqrt{(1+q)R_1/a}}\label{eq:mdot_mod}
\end{equation}
\citep{Marsh04}, where $R_1$ is the accretor radius and $a$ is the binary separation. This extreme case predicts a deviation from linearity of $0.23$ or $0.32$ in phase over our baseline, assuming a fully degenerate donor and an accretor mass of $0.60$ or $0.45$\Msun. This is clearly inconsistent with our timings, suggesting that a substantial fraction of the angular momentum is indeed fed back into the orbit.

\subsubsection{Period evolution with no mass transfer}
\label{sec:gr}

We consider also the scenario in which the donor star is detached from its Roche lobe. In this case the period will decrease as the components spiral in as angular momentum is lost due to gravitational radiation. The three curved tracks in the bottom panel of Figure \ref{fig:timings_mdot} show the predicted departure from a linear ephemeris for the detached case, for the same combinations of $M_1$ and $M_2$ used in the top panel. All three lines are a poorer fit to our timings compared to the corresponding semi-detached scenarios, as expected given the evidence of accretion.

\section{CONCLUSIONS}
\label{sec:conclusions}
We have obtained high time resolution lightcurves of the AM CVn star ES Cet with WHT+ULTRACAM and other smaller telescopes around the world. We observed the previously reported cyclical variability which has been confirmed to be orbital in origin. Our observations cover a baseline of $9$ years and we fit the lightcurves to obtain precise orbital timings over that period. We find that there is no evidence for a deviation from a linear ephemeris over this time period, although scatter in the timings is high due to flickering and large cycle-to-cycle variations in the light curve shape. There may be an underlying trend which is masked by this scatter. However, many predictions do suggest a lengthening orbital period which should be clearly detectable over this noise. Our findings suggest a low mass for the donor star, close or equal to the zero temperature mass of $0.062$\Msun. This would make the double white dwarf formation channel the most likely scenario for this system. Alternatively, the accretion rate in this system may be significantly below the long-term, secular rate for an accreting AM CVn star. ES Cet is the shortest period AM CVn star and the only ultracompact binaries with shorter periods are HM Cnc and V407 Vul, both of which have been shown to have a decreasing period. ES Cet may be in an intermediate phase between these systems and the longer period AM CVns.

\section*{ACKNOWLEDGEMENTS}
CMC and TRM are supported under grant ST/F002599/1 from the Science and Technology Facilities Council (STFC). ULTRACAM, SPL and VSD are supported by STFC grants PP/D002370/1 and PP/E001777/1. DS acknowledges the support of an STFC Advanced Fellowship. PAW and BW are supported by the National Research Foundation and the University of Cape Town. The results presented in this paper are based on observations made with the William Herschel Telescope operated on the island of La Palma by the Isaac Newton Group in the Spanish Observatorio del Roque de los Muchachos of the Institutions de Astrofisica de Canarias. This research has made use of NASA's Astrophysics Data System Bibliographic Services and the SIMBAD data base, operated at CDS, Strasbourg, France. We thank Lars Bildsten for helpful discussions.

\bibliography{escet}

\end{document}